\documentclass{aa}
\usepackage{psfig}

\begin{document}

\title{Simultaneous single-pulse observations of radio pulsars}
\subtitle{IV. Flux density spectra of individual pulses}
\author{M. Kramer
	\inst{1}
	\and
	A. Karastergiou 
	\inst{2,3}
	\and
        Y.~Gupta\inst{4}
	\and 
	S.~Johnston\inst{3}
        \and
        N.D.R.~Bhat\inst{5}
	\and
	A.G.~Lyne\inst{1}
}

\institute{Jodrell Bank Observatory, University of Manchester, 
 Macclesfield, Chesire SK11 9DL, UK
\and
Max-Planck Institut f\"ur Radioastronomie, Auf dem H\"ugel 69, 53121 Bonn, Germany
\and School of Physics, University of Sydney, NSW 2006, Australia
\and NCRA, TIFR, Pune University Campus, Ganeshkind, Pune 411007, India
\and Arecibo Observatory, HC3 Box 53995, Arecibo, Puerto Rico, PR 00612, USA
}

\abstract{ In this paper we demonstrate that a large, unexplored
reservoir of information about pulsar emission exists, that is
directly linked to the radiating particles and their radiation
process:  We present a study of flux density measurements of
individual pulses simultaneously observed at four different
frequencies.  Correcting for effects caused by the interstellar
medium, we derive intrinsic flux density spectra of individual radio
pulses observed at several frequencies for the first time.  Pulsar
B0329+54 was observed at 238, 626, 1412 and 4850 MHz, while
observations of PSR B1133+16 were made at 341, 626, 1412 and 4850
MHz. We derive intrinsic pulse-to-pulse modulation indices which show
a minimum around 1 GHz. Correlations between the flux densities of
different frequency pairs worsen as the frequency separation widens
and also tend to be worse for outer profile components. The single
pulse spectra of PSR B0329+54 resemble the spectra of the integrated
profile. However, the spectral index distributions for the single
pulses of PSR B1133+16 show significant deviations from a Gaussian.
This asymmetry is caused by very strong pulses with flux
densities exceeding the mean value by more than a factor of ten.  These
strong pulses occur preferentially at the trailing edge of the leading
component and appear to be broadband in most cases. Their properties
are similar to those of so-called giant pulses, suggesting that these
phenomena are related.
 }

\maketitle

\section{Introduction}\label{intro}

This work is the fourth part in a series describing simultaneous
multi-frequency observations of radio pulsars. The project had its
beginning in a collaboration known as the European Pulsar Network
(EPN) but has grown beyond its initial participating partners (Lorimer et
al.~ \cite{ljs98}) and now also utilizes telescopes outside Europe
such as the Giant Metre Radio Telescope (GMRT), India, which provided
low frequency coverage for the study presented here. The project's
observations are aimed at providing information and constraints on the
yet to be determined emission mechanism of radio pulsars.

While the properties of average profiles are determined by geometrical
factors to a large extent, the study of single pulses promises to be
the best way to investigate emission properties produced by the
radiation process.  The comparison of data obtained at several
frequencies at the same time offers the possibility of distinguishing
between propagation effects and of properties inherited from the
original radiation process. This has been demonstrated in the first
three paper of this series by Karastergiou et al.~(2001, 2002,
2003)\nocite{khk+01,kkj+02, kjk03}, where we concentrated on the
polarization properties of individual pulses.  In this paper, we shift
our focus towards the flux density spectra of single pulses and their
components.

Detailed knowledge about the spectral behaviour of individual emission
entities can help to put constraints on theoretical models and the
energy distribution of the emitting particles. Generally, only the
spectral behaviour of average pulse profiles is well known, but
obviously all information about dynamical processes in the pulsar
magnetosphere is lost in the averaging process. Hence, it is desirable
to study the radio spectrum of the single pulses themselves.

Simultaneous multi-frequency observations are difficult to realize, in
particular if total power information is to be obtained with a single
radio telescope. This emphasizes the need for coordinated
multi-station experiments such as reported here. In the past, rather
few such studies have appeared in the literature (see Karastergiou
et al.~2001).  Most
of the studies on simultaneous flux density measurements were done in
the early days of pulsar research. These were summarized by Backer \&
Fisher (1974)\nocite{bf74}, who provided the largest frequency
coverage so far from 250 to 8085 MHz, although they were limited to
single-polarization information for most of their frequencies.
However, in contrast to Robinson et al.~(1968)\nocite{rcg+68}, who
covered a wide range from 85 to 1410 MHz for PSR B1919+21, Backer \&
Fisher (1974)\nocite{bf74} did not present the spectra of single
pulses but averaged over a certain time in order to combat the effects
of interstellar scintillation (ISS).  Robinson et al.~did not correct
for ISS, while we know today that in the strong scattering regime
(i.e. at observing frequencies below a few GHz for low dispersion
measure pulsars), diffractive scintillation results in strong
modulations of the observed signal. ISS is therefore an important
problem to overcome when trying to derive the intrinsic radio
spectrum, and hence we pay particular attention to this aspect before
deriving and discussing the flux density spectra of individual radio
pulses. The plan for the rest of the paper is as follows: after
briefly summarizing the observations of two pulsars, PSRs B0329+54 and
B1133+16, we inspect the flux densities measured and correct
for the effects
of ISS. We then derive the individual flux density spectra and discuss
the results in the last section.

\section{Observations}

The data presented here were obtained on 2000 January 4, using the
Effelsberg 100-m telescope at 4850 MHz, the Lovell 76-m telescope at
Jodrell Bank at 1412 MHz and the GMRT at 626 MHz. The GMRT also
observed PSR B0329+54 at 238 MHz and PSR B1133+16 at 341 MHz,
respectively. The observations at the high frequencies using the
Effelsberg and the Lovell telescope have been described in
Karastergiou et al.~(2001, 2002).
The bandwidths used at these frequencies were 500 MHz
(Effelsberg) and 32 MHz (Lovell). At the GMRT we used bandwidths of 16
MHz at each frequency. The central frequencies for the lowest
receiving band differed for the two sources observed, as these signals
were recorded by the same data logging system
%i.e.~pulses at both frequencies appear in the same data window. 
(i.e. ~signals from both frequencies were added together after square-law
detection, and the pulsar's dispersion delay was used to separate the
two data streams during off-line analysis).
The lower observing frequency was chosen such that the dispersion delay 
relative to the 626 MHz resulted in well separated pulses. Data from all 
telescopes were stored  for off-line processing.

Using observing parameters summarized in Table \ref{obstab}, we
observed 2513 pulses (30 min) of PSR B0329+54 and 4375 pulses (86 min)
of PSR B1133+16. All data telescopes were converted to a common EPN
format and time-aligned following the procedures detailed in 
Karastergiou et al.~(2001).
The effective resolutions of our time series were 711.0 $\mu$s for
PSR B0329+54 and 1159.8 $\mu$s for PSR B1133+16, respectively.
Some measured pulses, in particular at the low frequencies of the
GMRT, were affected by radio interference. Such pulses, 251 for PSR
B0329+54 and 41 for PSR B1133+16, were excluded from the subsequent
analysis.

In Effelsberg and Jodrell Bank the pulses were compared to a signal of
a calibrated noise diode, which itself was compared to the strength of
known flux calibrators observed during pointing observations before,
after and during the observations, respectively.  Flux densities of
these point sources, such as NGC7027 and 3C48, were obtained from Peng
et al.~(2000)\nocite{pkkw00} and Ott et al.~(1994)\nocite{owq+94},
resulting in an estimated uncertainty in the measurements of about
10\%.

At the GMRT, known flux calibrators -- such as 3C147, 3C295 and 3C286
-- were observed before and after every pulsar observation.  Unlike
the pulsar observations, the calibration sources were observed
separately at two frequencies, i.e. ~one frequency at a time, using
the summed signal from only the antennas at the selected frequency.
From these observations, an effective calibration scale 
was computed for each of the two frequencies.  Using
the nearest calibration source observation, this scale was applied to
the ``on-off'' deflection of the pulsar, separately for each frequency
of observation, in order to flux calibrate the pulsar signal at
both frequencies.

As we will demonstrate later, the corresponding flux densities of the
average profiles are consistent with what is known from the
literature, giving us confidence in the accuracy of our single pulse
measurements.

\begin{table*}
\begin{center}
\caption{\label{obstab}
Observing and estimated scintillation parameters for both sources
presented in this paper. Observing frequency, $\nu$, and 
bandwidth, $\Delta \nu$, and observing times
are given in Cols.~3, 4 and 5, while expected ISS
de-correlation bandwidths and timescales are given in Cols.~5 and 6.
}
\begin{tabular}{llrrcccl}
\hline
\hline
PSR & Telescope & Frequency & Bandwidth & $t_{\rm obs}$ & $\Delta \nu_{\rm ISS}$ &
$\Delta t_{\rm ISS}$ & ISS Type\\
    &       & \multicolumn{1}{c}{(MHz)} & \multicolumn{1}{c}{(MHz)} &
\multicolumn{1}{c}{(min)} & \multicolumn{1}{c}{(MHz)} &
 \multicolumn{1}{c}{(min)} &  \\
\hline
B0329+54 & GMRT & 238 & 16 & 30 & 0.004 & 2 & strong\\
         & GMRT & 626 & 16 & & 0.4 &  5 & strong\\
         & Lovell & 1412 & 32 & & 17 & 13 & strong \\
         & Effelsberg & 4850 & 500 & & --  & 80 & weak \\
B1133+16 & GMRT & 341 & 16 & 86 & 1.0 & 3 & strong \\
         & GMRT & 626 & 16 & & 11.4 & 6 & strong \\
         & Lovell & 1412 & 32 & & 270 & 16 & strong \\
         & Effelsberg & 4850 & 295 & & -- & 14 & weak  \\
\hline
\end{tabular}
\end{center}
\end{table*}

\section{Interstellar Scintillation (ISS)}\label{iss}

\subsection{Estimated and observed parameters}

Before we can use flux density data for a computation of the
radio spectrum, we must study the possible impact of ISS. In
particular, we have to estimate the influence of diffractive
scintillation which occurs on timescales of minutes. Up to
frequencies of about 1 GHz, the sources are usually in the strong
scintillation regime, and the de-correlation bandwidth scales
as 
\begin{equation}
\Delta \nu_{\rm ISS} \propto \nu^{\beta}
\end{equation}
(e.g.~Cordes \& Rickett 1998)\nocite{cr98} where 
$\beta=4.4$ for a Kolmogorov turbulence spectrum. The modulation
timescale, $\Delta t_{\rm ISS}$ is expected to increase with frequency
as $\Delta t_{\rm ISS} \propto \nu^{1.2}$
(e.g.~Gupta 1995, Cordes \& Rickett 1998).

If the scintillations in time and in frequency are sufficiently well
sampled then the modulation index, $m$, defined as
\begin{equation}
\label{modidzeqn}
m^2 = \frac{<\!(S - <\!S\!>)^2 \!>}{<\!S\!>^2},
\end{equation}
is unity in the strong (diffractive) scintillation regime
if $S$ is the measured flux density and $<\!S\!>$ its mean value.
However, if the scintillation bandwidth is significantly less than
the observing bandwidth, the scintles will `wash out' and the measured
modulation index will be much lower.

In the weak scintillation regime, modulation decreases while
timescales are expected to decrease with frequency as well, $\Delta
t_{\rm ISS} \propto \nu^{-0.5}$.  For our sources, scintillation
parameters have been measured by Malofeev et al.~(1996)
\nocite{mss+96} and both sources are found to be in the weak
scintillation regime at 4850 MHz, with modulation indices measured by
Malofeev et al.~of only $m\sim0.12$ for B0329+54 and $m\sim0.22$ for
PSR B1133+16. Timescales expected for weak scintillation from the
ratio of Fresnel scale and velocity for both our pulsars are listed in
Table~\ref{obstab}.

\begin{figure}[t]
%\resizebox{\hsize}{!}{\includegraphics{flux0329.ps}}

\psfig{file=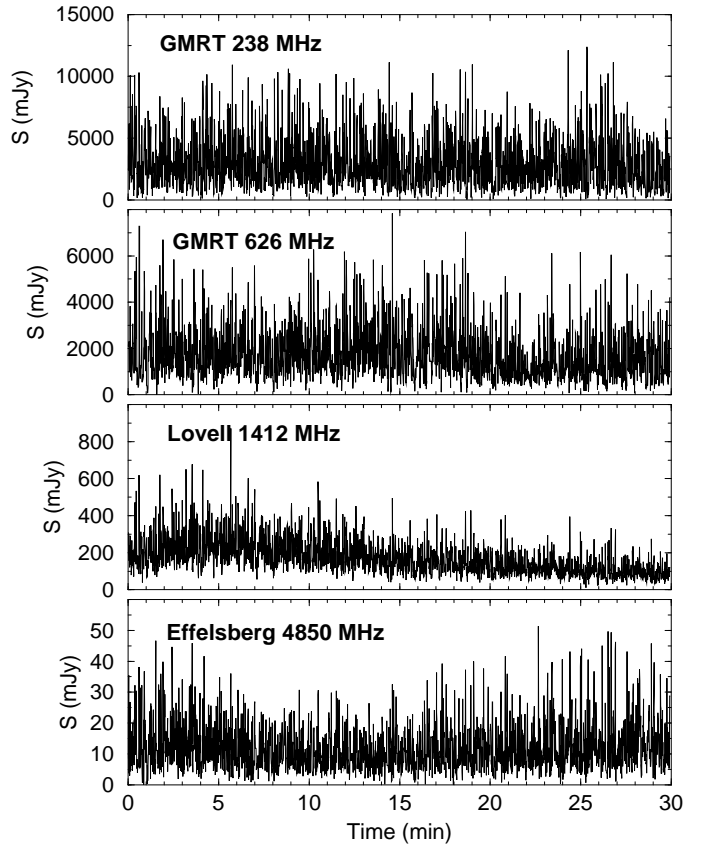,width=9cm}

\caption
{Equivalent continuum flux densities simultaneously observed for 
the full profile of PSR B0329+54.
Effects of ISS are clearly visible at 1412 MHz and may also
be present at 626 and 4850 MHz.
\label{flux0329}}
\end{figure}

Pulsar B0329+54's critical frequency, $\nu_c$, marking the transition
from the strong to weak scintillation regime, is estimated to be
around 3 GHz, consistent with the results by Malofeev et al.~(1996).
This places PSR B0329+54 in the
strong ISS regime at all our frequencies below 4850 MHz. 
For these frequencies, we estimate
the expected scintillation parameters by using measurements made by
Bhat et al.~(1999b) at 327 MHz, and Kondratiev et al.~(2001)
and references therein. We fit a power laws to data available
in a frequency range from 102 MHz to 1420 MHz and derive the
values listed in Table~\ref{obstab}. 
%
% and a frequency scaling of $\Delta
%\nu_{ISS} \propto \nu^{4.4}$ and $\Delta t_{ISS}\propto \nu^{1.2}$. 
%This
%indeed suggests a transition frequency of $\nu_c \sim 3.5$ GHz for PSR
%B0329+54.  
At 238 and 626 MHz, our observing bandwidths are much
larger than the expected de-correlation bandwidths, as displayed in
Table \ref{obstab}.  
The observing bandwidth at 1412 MHz is about
twice the estimated de-correlation bandwidth, so that some flux
density modulation can be
expected.
Indeed, all these expectations are confirmed in Fig.~\ref{flux0329} 
where we show the measured equivalent
continuum flux density for all four
frequencies. We will discuss details in Sect.~3.2.

\begin{figure}[t]

%\resizebox{\hsize}{!}{\includegraphics{flux1133.ps}}
\psfig{file=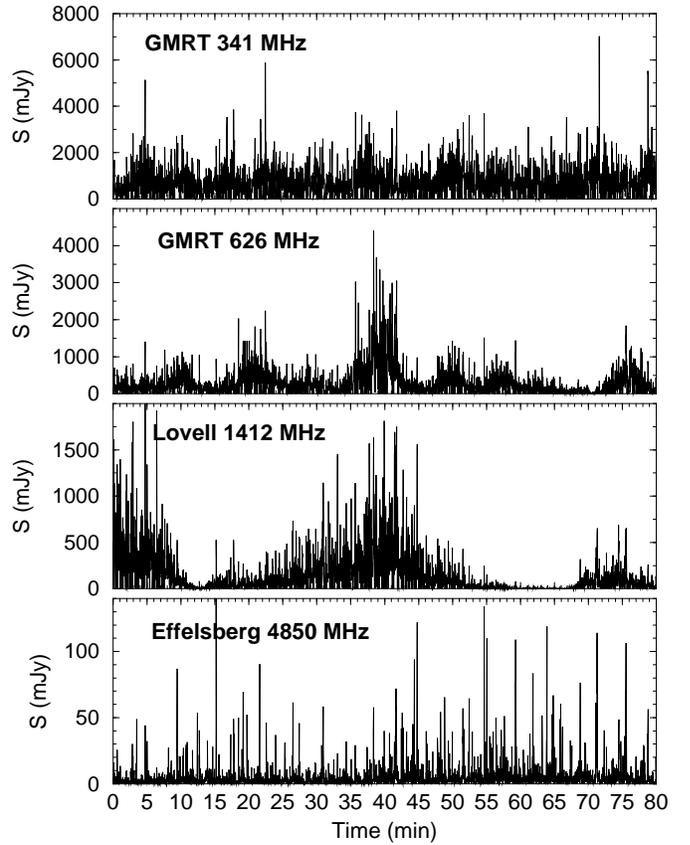,width=9cm}

\caption
{Equivalent continuum flux densities simultaneously observed for 
the full profile of PSR B1133+16.
Severe effects of ISS are present in particular at 626 and 1412
MHz.
\label{flux1133}}
\end{figure}

In the case of PSR B1133+16, Malofeev et al.~(1996) extrapolated the
critical frequency $\nu_c$ to a value of around 700 MHz. However,
inspecting Fig.~\ref{flux1133} we realize that this pulsar is
clearly in the strong scintillation regime at 1412 MHz.  Bhat et
al.~(1999b) measured three different scintillation bandwidths at a
frequency of 327 MHz.  
Using the mean value of those and a frequency
scaling of $\Delta \nu_{\rm ISS} \propto \nu^{4.4}$ and 
$\Delta t_{\rm ISS}\propto \nu^{1.2}$,
we estimate the transition frequency to
be about $\nu_c \sim 2$ GHz instead, with scintillation parameters as
listed in Table~\ref{obstab}. 
 At 341 MHz we expect to sample a number
of scintles, resulting in only little modulation. In contrast, at the
other frequencies the observing bandwidth is of same size or much
smaller than the de-correlation bandwidth. Strong intensity variation
should be expected on a timescale of 6 min at 626 MHz and 16 min at
1412 MHz which seems, in particular when considering the involved
uncertainties, to be in good agreement with the observations
(Fig.~\ref{flux1133}).  In the following we present a procedure
which is aimed at separating flux density variations of intrinsic
origin and those caused by ISS.

We note that in addition to diffractive scintillation, a refractive
branch exists as well (Sieber 1982)\nocite{sie82}.  Timescales
associated with refractive scintillation are much longer while the
involved modulations are much weaker (e.g.~Rickett
1990).\nocite{ric90} We can try to estimate the expected refractive
timescale in strong scintillation regime by using our derived
diffractive scintillation parameters and the relation
\begin{equation}
\Delta t_{\rm RISS} \sim u^2 \Delta t_{\rm ISS} 
\sim\frac{\nu}{\Delta \nu_{\rm ISS}}\times \Delta t_{\rm ISS}
\end{equation}
where $u$ is the strength of scattering given by the ratio of
Fresnel and coherence scale
(e.g.~Rickett 1990, 
Stinebring \& Condon 1990). \nocite{sc90a} For $\nu = 1412$ MHz,
we derive for PSR B0329+54, 
$\Delta t_{\rm RISS} \sim 830$ min, being much longer
than our total observing time.  For this pulsar, Stinebring et
al.~(1996) \nocite{sfm96} studied the refractive scintillation
properties in detail, observing flux density variations at 610 MHz of
about 30\%. This is consistent with the results obtained by Stinebring
et al.~(2000)\nocite{ssh+00} and those by Bhat et al.~(1999a)
at 327 MHz who also obtained measurements for PSR B1133+16. For
the latter pulsar, we estimate $\Delta t_{\rm RISS} \sim 80$ min at 1412 MHz
which is
the same as our observing time for this pulsar. Based on the results
by Stinebring et al.~(1996, 2000), Bhat et al.~(1999b)
% they have a modulation index of 0.2 at 327 MHz!
and an expected frequency scaling of $m_{\rm RISS}\propto \nu^{\sim 0.56}$
(e.g.~Stinebring et al.~2000), we expect modulation
indices due to RISS for both pulsars of $m_{\rm RISS}\la 0.4$ at 1410 MHz.  We
note that even such moderate values can lead to fairly large
deviations of the flux from the mean value at some individual epochs,
as the refractive flux density fluctuations usually do not show simple
Gaussian-like distributions. However, lacking information for epochs
adjacent to our observing period, we take the value of
$m_{\rm RISS}$ as a typical 1-sigma fractional error bar. Under the
assumption that we can successfully correct for
diffractive effects (see next Section), we hence adopt
uncertainty estimates for our flux density measurements
of 22\% (238, 341, 4850 MHz), 30\% (626 MHz) and 40\% (1412 MHz),
reflecting RISS and possible calibration errors.

\subsection{Isolating the effects of ISS}

In order to correct for the effects of scintillation and to obtain the
intrinsic flux density values, published flux density spectra are
usually derived from the mean values of a large number of independent
observations spread over a long time scale, i.e.~years. It is indeed
quite reasonable to assume that this procedure will average out the
flux density variations due to diffractive and refractive ISS, 
resulting in the intrinsic flux density value at a given frequency
(e.g.~Stinebring et al.~2000)

Figures~\ref{flux0329} and \ref{flux1133} indicate that we typically
observe several diffractive modulation timescales during our
measurements. Hence, we should expect the average measured flux
densities to be consistent with the values published for a given
frequency within the uncertainties estimated above.

\begin{figure}[t]

%\resizebox{\hsize}{!}{\includegraphics{spectra_av_u.ps}}
\psfig{file=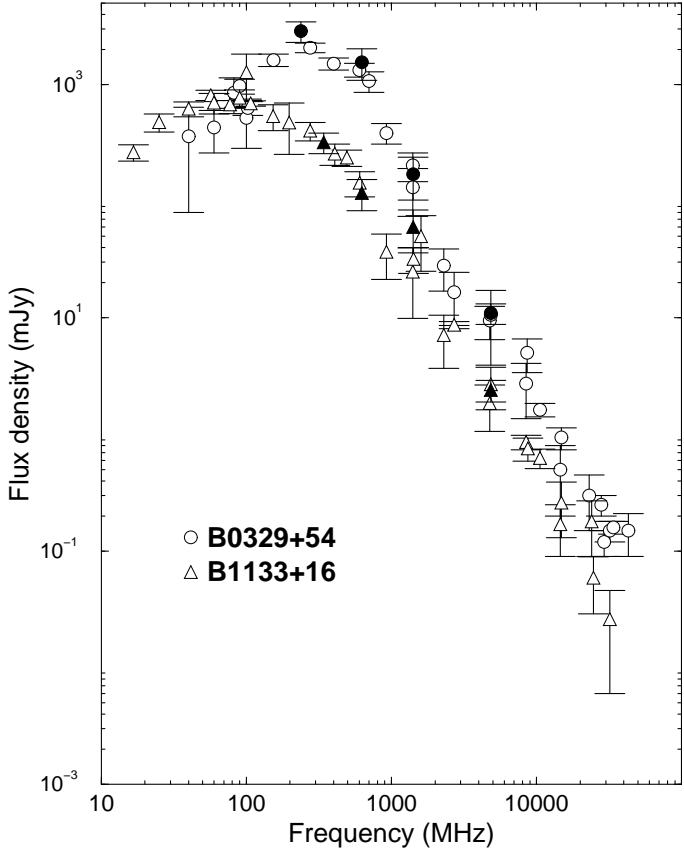,width=9cm}

\caption
{Spectra computed from equivalent continuum flux densities for B0329+54
(circles) and B1133+16 (triangles). The open symbols are
values taken from the literature, while the filled symbols
were derived from the single pulse measurements presented here.
\label{spectra}}
\end{figure}

The obtained flux densities are listed for both pulsars in
Table~\ref{fluxtab}.  Quoted uncertainties are derived from the error
of the mean and the earlier estimated uncertainties. All values are
plotted in Fig.~\ref{spectra} as filled symbols. The figure also
shows flux densities compiled by Maron et al.~(2000)\nocite{mkk+00}
and Malofeev \& Malov (1980)\nocite{mm80} as open symbols.  Indeed,
our average flux measurements agree very well with the published
values.

\begin{table}
\caption{\label{fluxtab}
Measured equivalent continuum flux densities as averages over the whole
observing time.
}
\centerline{
\begin{tabular}{rcc}
\hline
\hline
Frequency & B0329+54 & B1133+16 \\
\multicolumn{1}{c}{(MHz)} & S(mJy) & S(mJy) \\
\hline
238 & $2900\pm600$ & -- \\
341 & -- &     $320\pm60$ \\
626 & $1600\pm500$ & $120\pm40$\\
1412 & $170\pm70$ & $60\pm20$ \\
4850 & $11\pm2$ & $2.4\pm0.5
$ \\
\hline
\end{tabular}
}
\end{table}

While the average flux densities are consistent with the expected
values, we correct for short term variation due to ISS
according to the following
scheme: We compute a 200-sec running median for the time series which
we wish to correct. A time of 200 sec was chosen
to be longer than intrinsic pulse-to-pulse modulation and to be
smaller or of similar scale as the expected (diffractive
and refractive) scintillation time scales
(see Table \ref{obstab}). The flux density of each pulse is 
divided by this running median and the whole dataset is then rescaled
to be consistent with the initial average flux density.  According to
the scintillation parameters summarized in Table~\ref{obstab}, this
procedure is mostly important for the 1412 MHz times series of both
PSRs B0329+54 and B1133+16, as well as the 626 MHz time series of PSR
B1133+16. As a detail, we note that any missing pulses due to the
nulling phenomena (Backer 1970)
observed in PSR B1133+16, which will be studied
elsewhere (Bhat et al.~in prep.), were replaced by a preceding
detected pulse before computing the running median.

\begin{table}
\caption{\label{modidxtab}
Modulation indices for the variation in observed flux density
caused by 
interstellar scintillation, $m_{\rm ISS}$, and intrinsic
variation, $m_{\rm int}$
}
\centerline{
\begin{tabular}{rcccc}
\hline
\hline
Frequency & \multicolumn{2}{c}{B0329+54} & 
            \multicolumn{2}{c}{B1133+16} \\
\multicolumn{1}{c}{(MHz)} & $m_{\rm ISS}$ & $m_{\rm int}$ & $m_{\rm ISS}$ & $m_{\rm int}$ \\
\hline
238 & 0.08 & 0.71 & -- &  -- \\
341 & --   & --   & 0.16 & 0.95 \\
626 & 0.17 & 0.75 & 0.74 & 1.16 \\
1412 & 0.31 & 0.44 & 0.98 & 1.21 \\
4850 & 0.12 & 1.10 & 0.28 & 2.94 \\
\hline
\end{tabular}
}
\end{table}

We can use the time series computed from the 200-sec running median to
derive the modulation index defined in Eq.~\ref{modidzeqn} as a
result from intensity variation caused by interstellar scintillation,
$m_{\rm ISS}$. Similarly, the left-over, short-term variations are a
measure for the intrinsic pulse-to-pulse modulation, which we can
express by a modulation index computed from the corrected time series,
$m_{\rm int}$.  Both sets of values are listed in Table~\ref{modidxtab}.

While we would expect a modulation index of about unity in the strong
scintillation regime, the actual observed ISS modulation will depend on
the number of scintles sampled in time and frequency. In case of an
observing bandwidth large compared to the scintle bandwidth, we can
expect the modulation index to be reduced by a factor of approximately
$1/\sqrt{n}$ where $n$ is the number of scintles averaged
over. Indeed, our observations confirm this expected behaviour as can
be seen for instance, for PSR B0329+54 at 238 MHz. Our bandwidth for
this observation was about 1600 times the expected scintillation
bandwidth (see Table~\ref{obstab}), so that we would expect the
modulation index to be only of the order of 0.03
(Table~\ref{modidxtab}), which is close to the observed value after
applying our correction scheme.  For PSR B1133+16,
the observing bandwidth at 326 MHz is about 16 times larger than
the estimated scintillation bandwidth, and the expected modulation
index of $1/\sqrt{16}=0.25$ is again close to what is observed.

The observations also confirm our earlier expectations that the data
mostly affected by ISS are those at 1412 MHz and the PSR B1133+16 data
at 626 MHz. While our observing time for PSR B0329+54 is not
sufficient to fully sample the ISS variations at 1412 MHz in time,
resulting in a modulation index of somewhat less than unity, the
measured modulation index of PSR B1133+16 is indeed consistent with
unity at that frequency, as expected. At 626 MHz, we would expect a
modulation index of $1/\sqrt{16/11.4}=0.83$ (see Table~\ref{obstab})
which is close to the 0.74 measured for PSR B1133+16.  Moreover, both
pulsars are clearly in the weak scintillation regime at 4850 MHz, and
the derived modulation indices are consistent with those of Malofeev
at el.~(1996) obtained at the same frequency. We can therefore be
assured that our procedure can successfully separate the modulations
due to ISS from the intrinsic flux density variations.

\begin{figure}[t]

%\resizebox{\hsize}{!}{\includegraphics{modidx.ps}}
\psfig{file=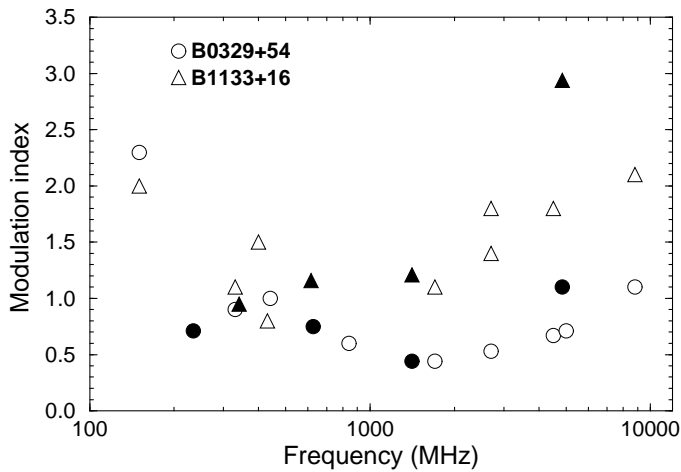,width=9cm}

\caption{\label{modidxfig}
Intrinsic pulse-to-pulse modulation indices, $m_{\rm int}$
as presented by Bartel et al.~(1980, open symbols) and
as derived from our measurements (filled symbols). The
peculiar frequency behaviour first pointed out by Bartel et al.~(1980)
is clearly visible.}
\end{figure}

After the correction scheme, which is equivalent to a high-pass
filtering, we are left with the intrinsic flux density time series.
Any possible imperfections of the scheme, such as
unmodelled refractive scintillation, are accounted for by adopting
uncertainties as derived in Sect.~3.1. We can therefore assume
that our modulation indices, $m_{\rm int}$, describe
intrinsic intensity variations. We can compare our $m_{\rm int}$
to those presented by Bartel et al.~(1980)\nocite{bsw80}. 
They studied the frequency dependence of the
intrinsic pulse-to-pulse modulation for a large sample of pulsars,
including PSRs B0329+54 and B1133+54. Interestingly, Bartel et
al.~found that $m_{\rm int}$ first decreases with frequency, before it
starts to rise again beyond a critical frequency, $\nu_{\rm m}^c$, which
they estimated to be $\nu_{\rm m}^c\sim1400$ MHz for PSR B0329+54 and
$\nu_{\rm m}^c\sim600$ MHz for PSR B1133+16, respectively. We show the
modulation indices presented by Bartel et al.~(1980) together with our
$m_{\rm int}$ measurements in Fig.~\ref{modidxfig}. Indeed, our values
agree well with the previous measurements, in particular when
considering that our data have been taken simultaneously at
different frequencies while those of Bartel et al.~were compiled
from different epochs. Moreover, both pulsars known to switch
radiation patters; PSR B0329+54 shows a very prominent moding behaviour
(e.g.~Bartel et al.~1982), while PSR B1133+16 is a  well known for
its nulling phenomenon. It is likely that these phenomena have
affected our analysis techniques and those of Bartel et al.~(1980)
differently, explaining some of the few differences. We note that our
very high modulation index for PSR B1133+16 is probably related to
another phenomenon, discussed in detail in Sect.~5.

Overall, it is worth emphasizing that the peculiar frequency behaviour
of the pulse-to-pulse modulation index is supported by our
measurements which were made simultaneously at the different
frequencies.

\section{Radio spectra of individual pulses}

\subsection{Correlations between frequencies}

\begin{figure}[t]

%\resizebox{\hsize}{!}{\includegraphics{correl0329.ps}}
\psfig{file=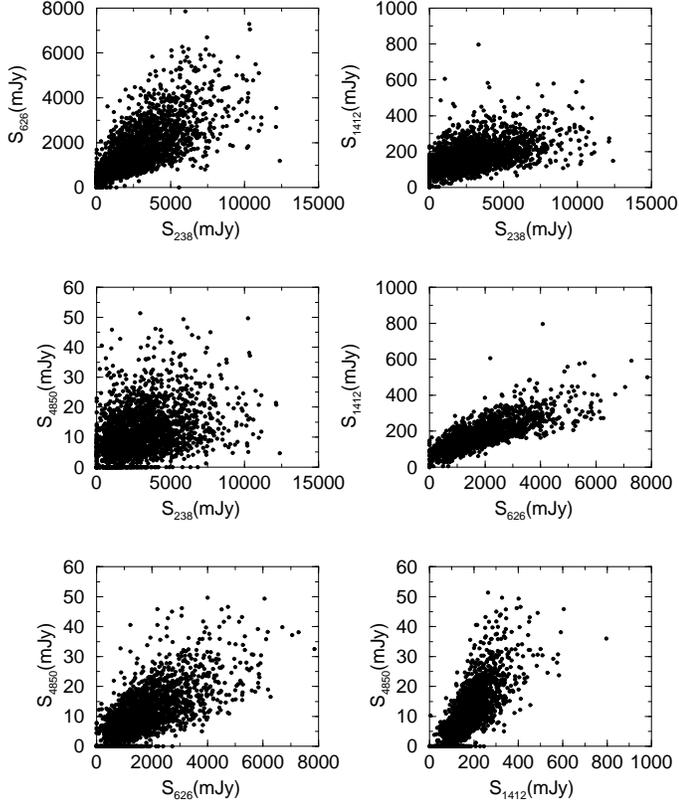,width=9cm}

\caption{\label{correl0329}
Correlation between the equivalent continuum
flux densities observed for
PSR B0329+54 simultaneously at four frequencies.
}
\end{figure}

Using the ISS corrected flux densities, we can inspect the correlation
of the intrinsic intensities between the different
frequencies. Fig.~\ref{correl0329} shows the flux densities measured
at our four frequencies for PSR B0329+54.
Corresponding correlation coefficients are summarized in 
Table~\ref{correl0329tab}. In order to estimate uncertainties 
in the correlation coefficients, we followed the example
of Bhat (1998)\nocite{bha98} and used 
the ``boot-strap'' method (e.g.~Efron 1979\nocite{efr79},
Press et al.~1992\nocite{ptvf92}). We also produced simulated
pulses with varying signal-to-noise ratio to study the effect of
noise on the measured correlation coefficient (cf.~Kardashev et al.~1986). 
We find that for our signal strengths, the reduction of the
correlation coefficient is typically less than 5\%. Hence,
rather than trying to correct for this effect, a corresponding
error was added in quadrature to the boot-strap uncertainties,
resulting in the values shown in Table~\ref{correl0329tab}.
Interestingly, the correlations
are clearly tighter for adjacent frequency bands than for those widely
separated. In fact, the best correlation can be observed between 626
and 1412 MHz, while the worst obviously exists for the 238/4850 MHz
pair.

\begin{figure}[t]

%\resizebox{\hsize}{!}{\includegraphics{profile0329.ps}}
\psfig{file=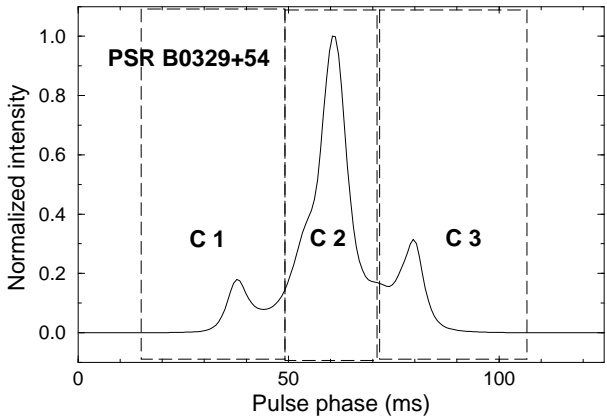,width=8cm}

\caption{\label{prof0329}
Average profile of PSR B0329+54 as observed at 1.41 GHz.
The three designated windows are referred to as Component 1, 2
and 3.
}
\end{figure}

The average profile of pulsars can be separated into individual
components that are well described by a Gaussian shape, for which a
spectral behaviour can be determined (e.g.~Kramer et al.~1994).
\nocite{kwj+94} Although this procedure should also be performed
ideally when studying components of single pulses, this is difficult
to achieve with single pulses which are not necessarily of Gaussian
shape.  In order to study the basic spectral behaviour of individual
single pulse components, we instead define windows across the pulse
for which we study the flux density separately. For PSR B0329+54 we
choose the three windows as shown for the 1.41-GHz profile in
Fig.~\ref{prof0329}. We are aware that this is only a crude
approximation, as it has been shown that the profile consists of at
least five (Kramer 1994) \nocite{kra94} or even nine components
(Gangadhara \& Gupta 2001)\nocite{gg01}, but we believe that more
complicated analysis will not alter the results significantly. We note
that the windows chosen are adequate for the profiles at all
frequencies.  The
correlation coefficients for these components separately are 
also given in Table~\ref{correl0329tab}.
Obviously, the best correlation
occurs for the central component, while the worst is present for
trailing one. 
We can compare these results to those of Kardashev 
et al.~(1986)\nocite{knn+86}, who looked 
at phase-resolved correlation coefficients
between widely-spaced frequencies.
Even though we can expect our results to differ somewhat
as we study only the correlation of the flux densities measured in
a given, much wider window, rather than attempting a phased-resolved analysis
as already done in the first paper of this series by Karastergiou
et al.~(2001), the results agree very well. Kardashev et al.~studied
pulses observed at 102.5 MHz and 1700 MHz and find correlation 
coefficients of about 0.38 and 0.43 for longitudes roughly situated within
our windows C2 and C3. Given their much lower frequency of 102.5 MHz,
their obtained values are in good agreement. It is difficult
to judge the observation 
that their C3 value is larger than their value determined
for C2, as they do not estimate uncertainties and as their
C3 value is in fact smaller before a noise-correction is applied.

\begin{table}
\caption{\label{correl0329tab} 
Correlation coefficients for PSR B0329+54 as determined
from the equivalent continuum flux densities measured between frequency
pairs for the full pulse and the
components 1, 2 and 3. Uncertainties have been computed using
the boot-strap method and simulated pulses (see text for details).
}

\begin{tabular}{rlr@{$\pm$}lr@{$\pm$}lr@{$\pm$}lr@{$\pm$}l}
\hline
\hline
\noalign{\smallskip}
Freq. & & \multicolumn{2}{c}{626 MHz} & \multicolumn{2}{c}{1412 MHz} & 
    \multicolumn{2}{c}{4850 MHz} \\
\noalign{\smallskip}
\hline 
\noalign{\smallskip}
238 MHz & Pulse & 0.67 & 0.03 & 0.48 & 0.03 & 0.31 & 0.03 \\
        & C1    & 0.60 & 0.04 & 0.47 & 0.04 & 0.15 & 0.03 \\
        & C2    & 0.69 & 0.04 & 0.54 & 0.03 & 0.39 & 0.03 \\
        & C3    & 0.43 & 0.03 & 0.29 & 0.02 & 0.26 & 0.02 \\
\noalign{\smallskip}
626 MHz & Pulse &   \multicolumn{2}{c}{ } & 0.79 & 0.04 & 0.66 & 0.04 \\
        & C1    &   \multicolumn{2}{c}{ } & 0.68 & 0.04 & 0.28 & 0.03 \\
        & C2    &   \multicolumn{2}{c}{ } & 0.83 & 0.04 & 0.72 & 0.04 \\
        & C3    &   \multicolumn{2}{c}{ } & 0.58 & 0.04 & 0.22 & 0.02 \\
\noalign{\smallskip}
1412 MHz & Pulse& \multicolumn{2}{c}{ } & \multicolumn{2}{c}{ } & 0.78 & 0.04 \\
         & C1   & \multicolumn{2}{c}{ } & \multicolumn{2}{c}{ } & 0.51 & 0.04 \\
         & C2   & \multicolumn{2}{c}{ } & \multicolumn{2}{c}{ } & 0.79 & 0.04 \\
         & C3   & \multicolumn{2}{c}{ } & \multicolumn{2}{c}{ } & 0.60 & 0.04 \\
\noalign{\smallskip}
\hline
\end{tabular}
\end{table}

\begin{figure}[t]

%\resizebox{\hsize}{!}{\includegraphics{correl1133.ps}}
\psfig{file=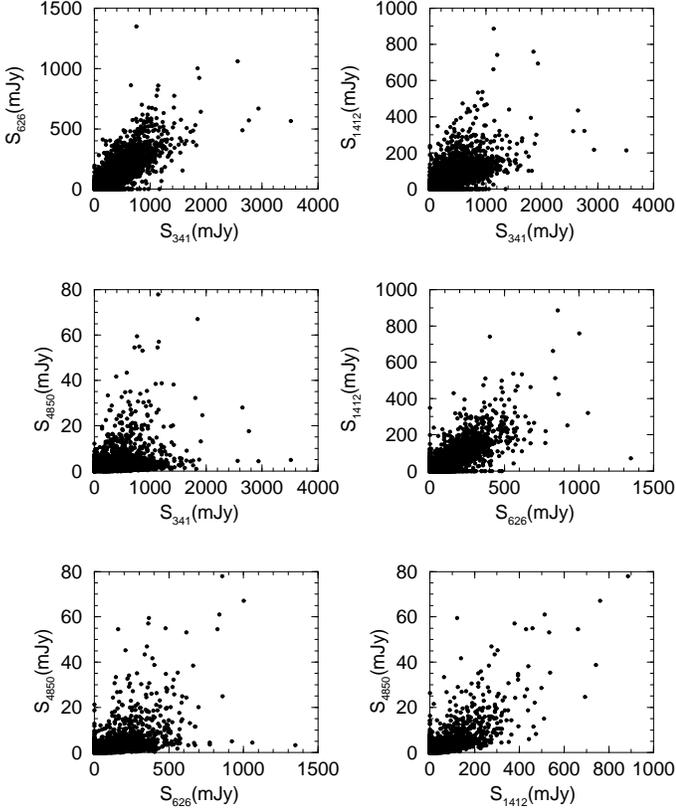,width=9cm}

\caption{\label{correl1133}
Correlation between the equivalent continuum flux densities observed for
PSR B1133+16 simultaneously at four frequencies.
}
\end{figure}

\begin{figure}[t]

%\resizebox{\hsize}{!}{\includegraphics{profile1133.ps}}

\psfig{file=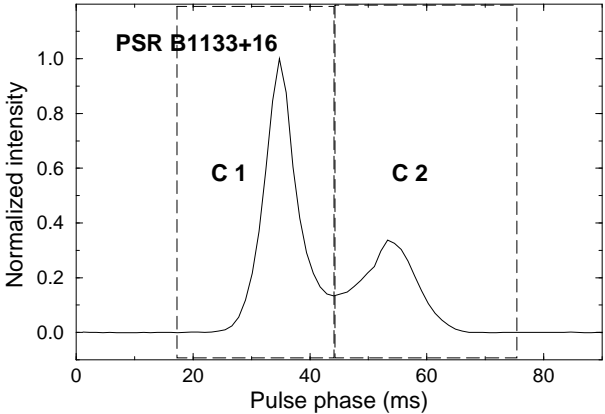,width=8cm}

\caption{\label{prof1133}
Average profile of PSR B1133+16 as observed at 1.41 GHz.
The two designated windows are referred to as Component 1 and 2.
}
\end{figure}

The data obtained for PSR B1133+16 are shown in
Fig.~\ref{correl1133} 
whereas correlation coefficients computed as
for PSR B0329+54 are summarized in Table~\ref{correl1133tab}. 
 Again
the least correlation is observed for the frequency pair with the
widest separation.  In order to study the behaviour of profile
components, we again define pulse windows as shown for PSR B1133+16 in
Fig.~\ref{prof1133}. These two components are also only an
approximation, since Kramer (1994) showed that the central bridge
emission should be modelled as a separate component, perhaps
representing the graze of an inner cone in a nested cone structure.
Generally, the correlation coefficients for the two profile components
(again in brackets) are larger than that for the full profile. The
only exception is for the correlation between 1412 MHz and 4850 MHz,
where at both frequencies the leading component is heavily
dominating. 
Again, our results are consistent with those by Kardashev et al.~(1986)
who also find the first component to be better correlated than the
second. Manchester et al.~(1989)\nocite{mps89} looked at this pulsar
at three different frequency pairs, i.e.~410--928 MHz, 410--5860 MHz
and 928--5600 MHz, respectively. As Kardashev et al.~(1986), they
performed a phase-resolved correlation analysis, finding again
a better correlation for the first component. While a proper comparison
is difficult due to the difference in freqencies and the lack
of estimated uncertainties, their C1 appears to show the least
correlation between 928 and 5600 MHz, while C2 exhibits the
lowest correlation coefficient for 410 and 5860 MHz.
Some of this difference is most likely due to a
frequency-selective nulling behaviour of this
pulsar.
We note that during our observations nulling did not always occur
simultaneously at all four frequencies.
As we will detail in Bhat et
al.,~(in prep.), while the majority of nulls are broadband, a
significant fraction of them occur only over a narrow frequency range.

\begin{table}
\caption{\label{correl1133tab} 
Correlation coefficients for PSR B1133+16 as determined
from the flux densities measured between frequency
pairs for the full pulse and the
components 1 and 2.
}

\begin{tabular}{rlr@{$\pm$}lr@{$\pm$}lr@{$\pm$}lr@{$\pm$}l}
\hline
\hline
\noalign{\smallskip}
Freq. & & \multicolumn{2}{c}{626 MHz} & \multicolumn{2}{c}{1412 MHz} & 
    \multicolumn{2}{c}{4850 MHz} \\
\noalign{\smallskip}
\hline 
\noalign{\smallskip}
341 MHz & Pulse & 0.72 & 0.04 & 0.51 & 0.03 & 0.31 & 0.02 \\
        & C1    & 0.77 & 0.04 & 0.57 & 0.03 & 0.35 & 0.04 \\
        & C2    & 0.76 & 0.05 & 0.61 & 0.04 & 0.32 & 0.03 \\
\noalign{\smallskip}
626 MHz & Pulse &   \multicolumn{2}{c}{ } & 0.73 & 0.04 & 0.49 & 0.04 \\
        & C1    &   \multicolumn{2}{c}{ } & 0.78 & 0.04 & 0.55 & 0.05 \\
        & C2    &   \multicolumn{2}{c}{ } & 0.74 & 0.05 & 0.48 & 0.04 \\
\noalign{\smallskip}
1412 MHz & Pulse& \multicolumn{2}{c}{ } & \multicolumn{2}{c}{ } & 0.74 & 0.04 \\
         & C1   & \multicolumn{2}{c}{ } & \multicolumn{2}{c}{ } & 0.76 & 0.05 \\
         & C2   & \multicolumn{2}{c}{ } & \multicolumn{2}{c}{ } & 0.69 & 0.04 \\
\noalign{\smallskip}
\hline
\end{tabular}
\end{table}

\subsection{Single pulse spectra}

\begin{figure}

%\resizebox{\hsize}{!}{\includegraphics{speccomb0329.ps}}
\psfig{file=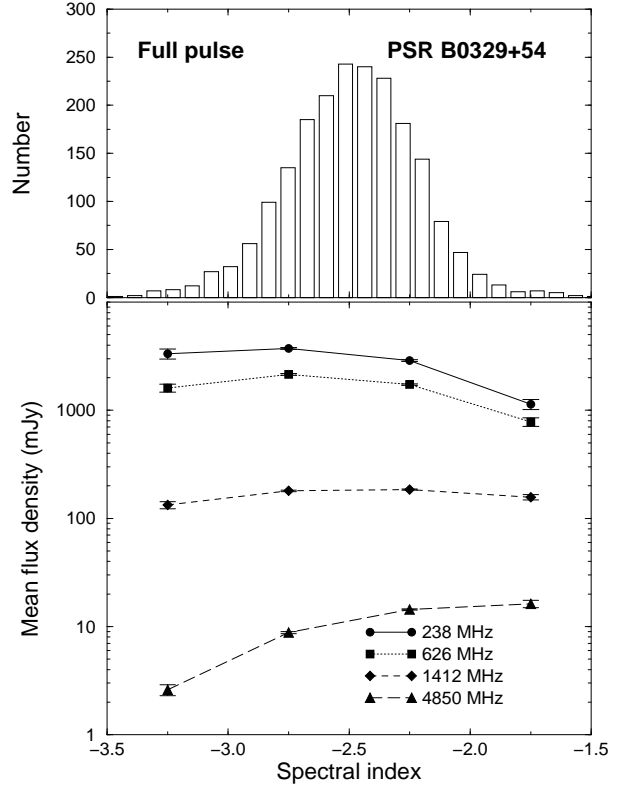,width=8cm}

\caption{\label{spec0329fig} 
{\em Top:} Distribution of spectral
indices obtained for single pulses of PSR B0329+54 using the full
pulse.  {\em Bottom:} Mean of the equivalent continuum 
flux density of pulses with a spectral
index within an interval of 0.5. 
%Average flux densities are plotted in the
%centre of the corresponding bin.  
}
\end{figure}

It is clear from Fig.~\ref{spectra} that the radio spectrum of PSR
B0329+54 exhibits a maximum at around 300 MHz, close to or even above
our lowest observing frequencies. In order to compare the single pulse
spectra to the average radio spectrum, we therefore will not include
this frequency in the process of fitting a power law spectrum
$S(\nu)\propto \nu^\alpha$ to the single pulse flux densities of this
pulsar. The distribution of
spectral indices using the full pulse window is presented in the top
panel of Fig.~\ref{spec0329fig}, showing a Gaussian shape. In order
to compute the mean spectral index from the individual fits to the
single pulses, we weight each result by the reduced
$\chi^2$-value obtained in each fit, 
hence greatly reducing the impact of possible bad
fits. We obtain a mean spectral index of $\bar{\alpha}=-2.4\pm0.1$,
which is consistent with the spectral indices obtained from using the
published flux densities in the same frequency range as well as using
the values obtained from the average profiles of our observations (see
Table~\ref{spec0329tab}). This result may be compared to the results
of a dual-frequency observation of PSR B0329+54 made at 327 MHz and
2695 MHz by Bartel \& Sieber (1978). Their spectral index distribution
also shows a Gaussian shape, even though their lowest frequency is
very close to the spectral turn-over. Due to the use of 327-MHz data, one
would expect that their mean spectral index may be biased towards
flatter spectra, but Bartel \& Sieber adjusted their distribution to
have the same median as the average pulse spectra, so that no detailed
comparison is possible. Additionally, we note that they did not
correct their data for ISS effects which they estimated to be small
for this pulsar at their frequencies.

\begin{figure}[t]

%\resizebox{\hsize}{!}{\includegraphics{speccomp0329.ps}}
\psfig{file=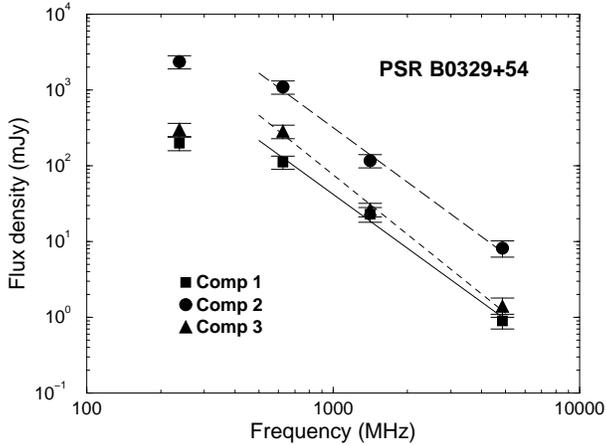,width=8cm}

\caption{\label{speccomp0329}
Component spectra for PSR B0329+54 using equivalent
continuum flux densities as measured of the average profiles
at each frequency.
}
\end{figure}

Computing the spectra for the components shown in
Fig.~\ref{prof0329} from the flux densities of the average pulse
profiles, we obtain the spectra shown in Fig.~\ref{speccomp0329}
where we again ignore the lowest frequency due to the presence of a
spectral turn-over.  In contrast to the general behaviour of pulsars
(e.g.~Rankin 1983, Lyne \& Manchester 1988), \nocite{lm88}
\nocite{ran83} the central component, Comp 2, does not exhibit the
steepest spectrum. Kramer (1994) obtained the same result
for the average profile using five Gaussian components in a more
sophisticated analysis in a frequency range from 1410 MHz to 10550
MHz. Our inner component corresponds roughly to his central three
components, for which we find an index of $-2.2\pm0.1$ using his
data. This is consistent with our result. However, while both his
leading outer components have a steeper spectrum than the trailing
ones, we observe the opposite with the steepest spectrum for C3.  It
is conceivable that the different frequency range covered is most
likely responsible, indicating a possible break in the component
spectra. 

\begin{figure}[t]
\psfig{file=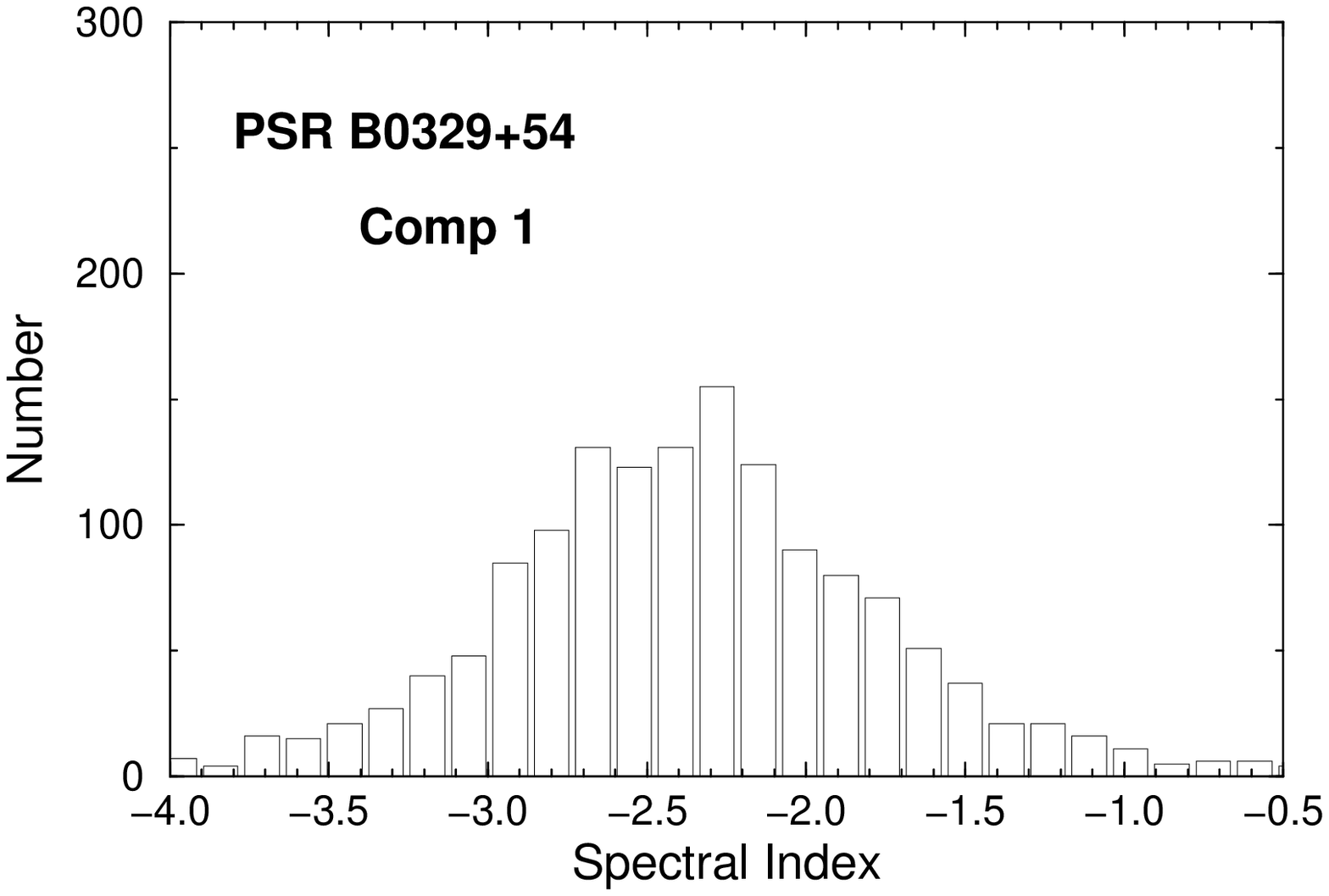,width=8cm}

\psfig{file=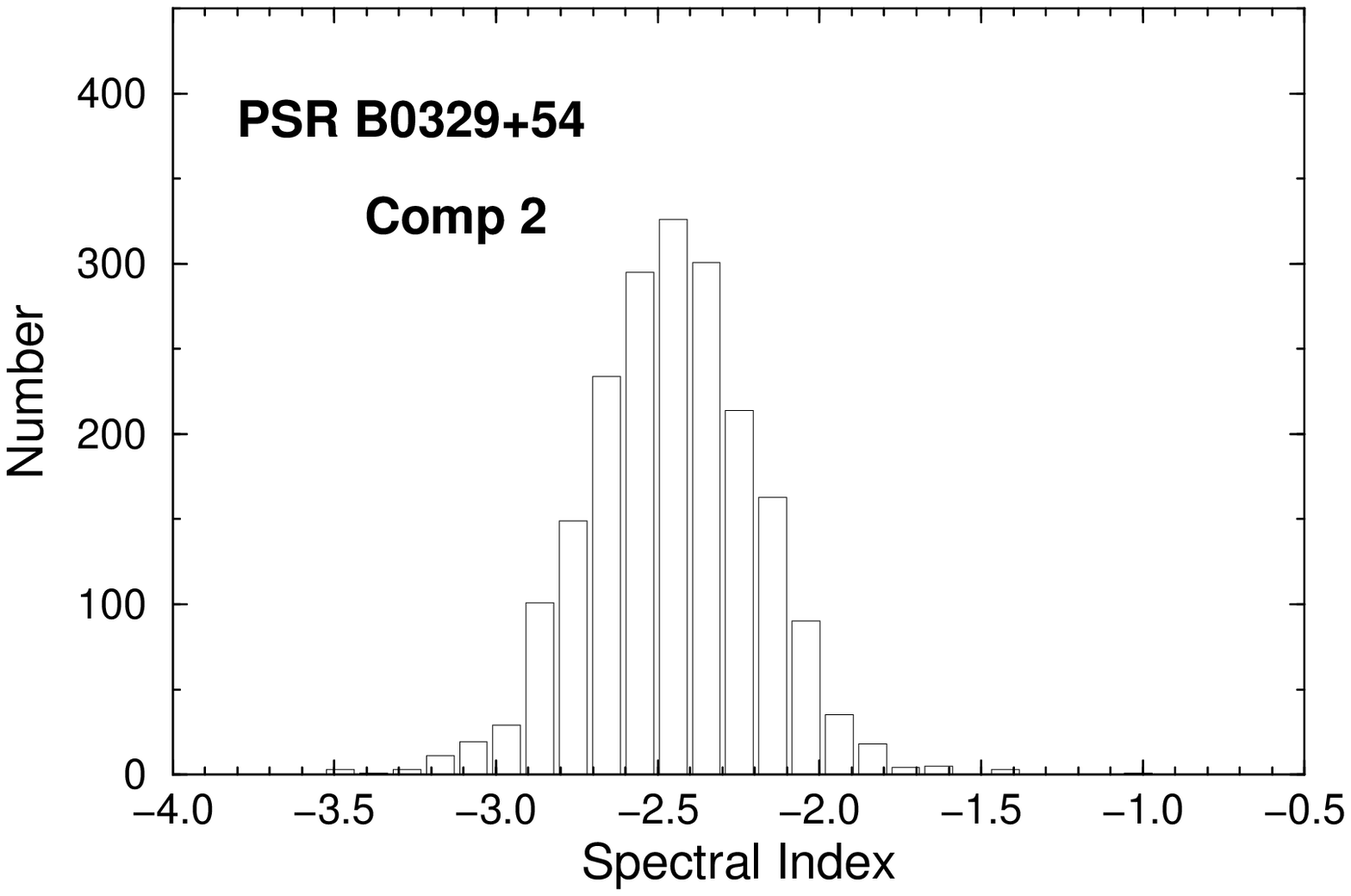,width=8cm}

\psfig{file=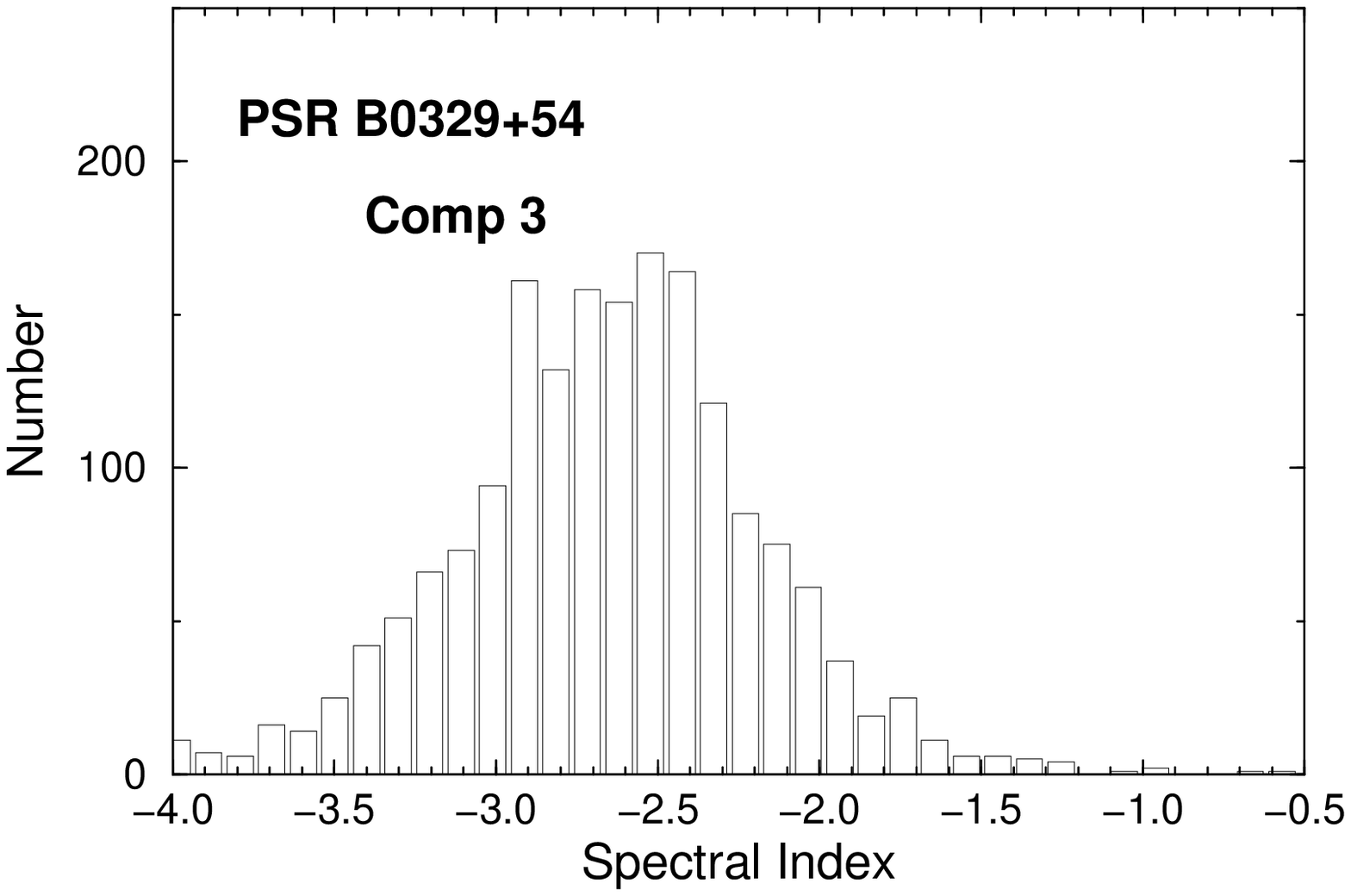,width=8cm}

%\resizebox{\hsize}{!}{\includegraphics{spec0329c1.ps}}
%\resizebox{\hsize}{!}{\includegraphics{spec0329c2.ps}}
%\resizebox{\hsize}{!}{\includegraphics{spec0329c3.ps}}

\caption{\label{spec0329cfig}
Distribution of spectral indices obtained for single pulses
of PSR B0329+54 using the three pulse windows shown in Figure
8.
}
\end{figure}

All obtained spectral indices are summarized in
Table~\ref{spec0329tab}. We also list the results obtained from the
$\chi^2$-weighted distributions of single pulse spectra which are
shown in Fig.~\ref{spec0329cfig}. The results are again
consistent. Not surprisingly, the distribution of the central,
dominating component shows the largest resemblance to the distribution
of the full pulse. The distributions for the outer components are much
wider. While these components are weaker, their spectra are still well
defined.  The mean errors of the spectral indices are similar for all
three components, and we do not find any correlation between the
goodness-of-fit, $\chi^2$, and the fitted spectral indices.  Hence, we
believe that the distributions are intrinsically wider. We will
revisit this observation later in light of the result for PSR B1133+16
described in the following.

\begin{table}
\caption{Spectral indices obtained for PSR B0329+54 
in the frequency range from 600 to 4850 MHz. The results
using the published flux densities are quoted in Col.~2,
using the flux densities from the average profile in Col.~3
and the weighted mean values of the single pulse spectra in Col.~4.
Corresponding medians are quoted in brackets.
\label{spec0329tab}}

\tabcolsep3pt
\begin{tabular}{lcccc}
\hline
\hline
 & Published & Average & \multicolumn{2}{c}{Single} \\
\hline
Full pulse & $-2.51\pm0.13$ & $-2.4\pm0.1$ & $-2.4\pm0.1$ & ($-2.44$)\\
Comp.~1 & -- &  $-2.4\pm0.2$ & $-2.2\pm0.1$ & ($-2.32$) \\
Comp.~2 & -- &  $-2.4\pm0.2$ & $-2.3\pm0.1$ & ($-2.40$)\\
Comp.~3 & -- &  $-2.6\pm0.2$ & $-2.5\pm0.1$ & ($-2.59$)\\
\hline
\end{tabular}
\end{table}

\begin{figure}[t]

%\resizebox{\hsize}{!}{\includegraphics{specindxfull1133.ps}}
\psfig{file=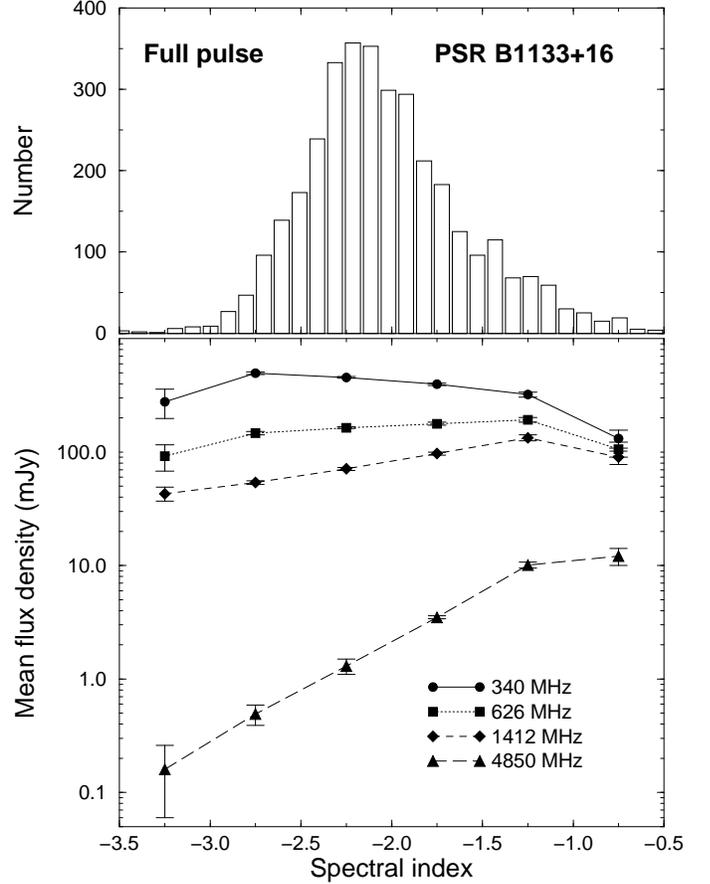,width=9cm}

\caption{\label{spec1133fig}
{\em Top:} Distribution of spectral
indices obtained for single pulses of PSR B1133+16 using the full
pulse.  {\em Bottom:} Mean of the equivalent continuum
flux density of pulses with a spectral
index within an interval of 0.5. 
%Average flux densities are plotted in the
%centre of the corresponding bin.  
}
\end{figure}

We repeat the same analysis for PSR B1133+16. Here all our observing
frequencies are obviously beyond the low frequency turn-over, so that
we perform a power law fit to all frequencies.  As we will detail in
Bhat et al.~(in prep.), nulling does not occur simultaneously at all
frequencies for this pulsar.  We fit a power law if at least three
frequencies show detections of pulses (or pulse components,
respectively). The distribution for the full pulse profile is shown in
the top panel of Fig.~\ref{spec1133fig}. The corresponding
($\chi^2$-weighted) mean spectral index (see Table~\ref{spec1133tab})
is somewhat steeper than that for the average profile. While
the values are just consistent within the errors, we note
that the single pulse distribution is not of Gaussian
shape, as in the case for PSR B0329+54, but that it shows a tail
towards flatter indices. Again, this is consistent with the results of
a dual-frequency observation by Bartel \& Sieber (1978) made at 327
MHz and 2695 MHz.  Their spectral index distribution is shifted
towards flatter spectra when compared to our results, but as mentioned
before Bartel \& Sieber adjusted their distribution to have the same
median as the average pulse spectra (known at that time).
Additionally, they did not correct for effects caused by ISS.  Whilst
no detailed comparison is therefore possible, the asymmetry in the
distribution is striking.

\begin{figure}[t]

%\resizebox{\hsize}{!}{\includegraphics{speccomp1133.ps}}
\psfig{file=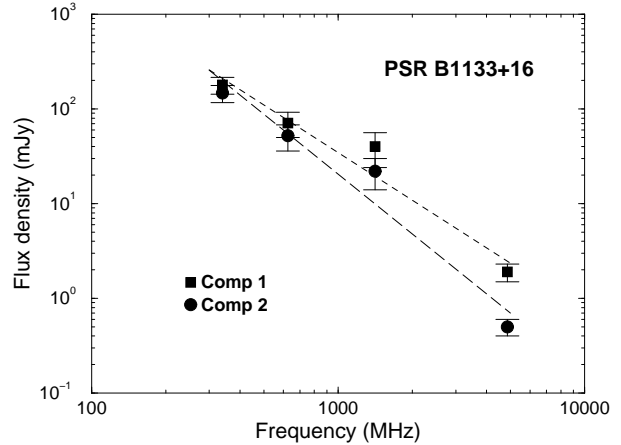,width=8cm}

\caption{\label{speccomp1133}
Component spectra for PSR B1133+16 using equivalent
continuum flux densities as measured of the average profiles
at each frequency.
}
\end{figure}

Using the average flux densities for the component windows (see
Fig.~\ref{prof1133}), we obtain the spectra shown in
Fig.~\ref{speccomp1133}. Corresponding spectral indices are listed
in Table~\ref{spec1133tab}. The values are in very good agreement with
the results obtained by Kramer (1994) using Gaussian components,
indicating again the validity of our relatively crude approach of
windowing.

The distributions of the spectral index obtained for each component
from the single pulses are shown in Fig.~\ref{speccomp1133} and
corresponding numbers are listed in Table~\ref{spec1133tab}.  It is
interesting to note that the width of the spectral index distributions
($\Delta \alpha \approx1$) is similar to that of the outer components of
PSR B0329+54. All of them are so-called ``conal'' components
(e.g.~Rankin 1983, Lyne \& Manchester 1988) which are viewed
further away from the magnetic meridian. It has been shown that
geometrical factors can severely affect the spectral index observed
for a given profile component (Kramer et al.~1994, Sieber 1997)
\nocite{kwj+94,sie97} and it seems therefore possible that geometry
can also affect the spectral index distribution of single
pulses. Indeed, it is conceivable that the the wider spectral index
distributions as well as the observed weaker correlations (between
pairs of frequencies) for PSR B0329+54's outer components (see
previous section) may be caused by the same underlying effect.  Future
comparisons of more single pulse spectra of ``core'' and ``conal''
components will help answer this question.

For the trailing component of PSR B1133+16, the mean of the spectral
index distribution is the same as for the average profile
spectrum. This is not the case for the dominating leading component or
the full pulse profile.  It is not obvious why this should be
expected, in particular as PSR B0329+54 behaved differently. We will
explore the possible reason for this effect in the following section.

\subsection{Spectral index vs. pulse strength}

An obvious difference in the spectral index distributions for the
single pulses of PSRs B0329+54 and B1133+16 is the asymmetry and
deviation from a Gaussian shape in the case of PSR
B1133+16. Constructing these histograms we lose information about the
relative strength of the observed single pulses, which is, together
with the relative occurrence of weak and strong pulses, nevertheless
relevant for the formation of the average profile. In order to recover
this information, we group the single pulses into intervals (bins) of
spectral indices, computing the average flux density at all
frequencies for a given spectral index bin.  We present the results in
the bottom panels of Figs.~\ref{spec0329fig} and \ref{spec1133fig}
where we show the mean flux density for a pulse of given spectral
index interval.  We note that their is often a wide scatter of flux
densities within a spectral index bin.  As a consequence, in few cases
when the number of pulses for a given bin is small, the spectral index
derived from these average flux densities shown in Figs.~\ref{spec0329fig} 
and \ref{spec1133fig} may not be identical to the
one of the associated spectral index bin.  However, the trend, as
confirmed by the computation of the medians of the flux densities, is
clear.

As one can see, for PSR B1133+16 there is a strong tendency for steep
spectrum pulses to be weaker than those with a flatter spectrum
(Fig.~\ref{spec1133fig}).  This trend is strongest for pulses at 4.85
GHz, while there is a less severe but still very significant
dependence at 626 MHz and 1412 MHz.  The curve at 340 MHz is
relatively flat with a trend into the opposite direction. The overall
trend only flattens out for those 5\% of pulses with a spectral index
greater than $-1$.

For PSR B0329+54 the behaviour is strikingly different.  While the
mean flux density only drops significantly at 4.85 GHz for those 2\% of
pulses with a spectral index steeper than $-3$, the curves are
otherwise remarkably flat at all frequencies, in particular for the
vast majority of the pulses in the centre of the spectral index
distribution.  This is even true for the lowest frequency of 238 MHz,
which follows the curves of the other frequencies, even though we have
seen that this frequency is below the spectral turn-over. This
observation may provide an important clue in identifying the origin of
the spectral turn-over. 

For PSR B1133+16, Fig.~\ref{spec1133fig} points to the existence of
pulses that are much stronger than the average at 4.85 GHz, but not
necessarily at the lowest frequencies, so that the overall spectrum of
these pulses is flatter than the average. This then causes an
asymmetry in the resulting spectral index distributions, as observed.

At high frequencies the pulse profile of PSR B1133+16 is dominated by
the leading component, as is seen in the spectral indices
of Table~\ref{spec1133tab}. At 10.55 GHz, the trailing component is
almost undetectable, and at 32 GHz the pulse profile is represented
by the leading component alone (Kramer et al.~1996).  Hence, we can
expect that the 4850 MHz data in Fig.~\ref{spec1133fig} are dominated
by the behaviour of component 1. Indeed comparing the mean flux
density for pulses with a spectral index $\alpha<-2.5$ to that of
pulses with $\alpha>-1.5$, we find $0.21\pm0.01$ and $8.7\pm0.5$ mJy,
respectively. For the second component, the difference is very much
smaller, i.e.~$0.51\pm0.01$ and $1.4\pm0.1$ respectively, but the
fact that there is a difference in these values at all helps in
understanding the asymmetry also observed in the spectral index
distribution for this component.

\begin{figure}[t]

%\resizebox{\hsize}{!}{\includegraphics{profall1133.ps}}
\psfig{file=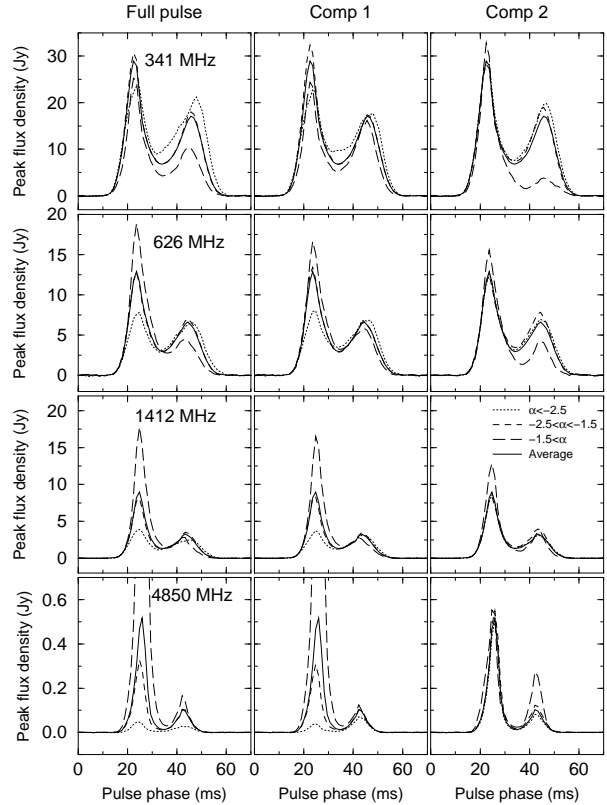,width=8cm}

\caption{\label{profall1133}
Average pulse profiles of PSR B1133+16 
computed for sets of pulses with
spectral indices measured in the intervals
$\alpha\le-2.5$,
$-2.5<\alpha\le-1.5$ and $\alpha>-1.5$, respectively,
for the whole profile and component 1 and 2.
The y-axis shows corresponding peak flux densities.
}
\end{figure}

It is the combined effect of pulse strength and relative occurrence
which produces the average profile at a given frequency. While we have
studied these two aspects independently from each other,
we can attempt to understand their relative importance by
computing the mean pulse profiles for each spectral index
bin. In order to facilitate the discussion we use coarse
spectral index bins of $\alpha\le-2.5$, $-2.5<\alpha\le-1.5$ and
$\alpha>-1.5$, respectively.  The results are shown in 
Fig.~\ref{profall1133} for all frequencies and for the spectral indices
measured for the full pulse and component 1 and 2.  This figure
indeed confirms our previous considerations: The figures for the full
pulse look very similar to those of component 1, as expected. At high
frequencies the flat spectrum profiles are dominated by a very strong
first component, while this effect gradually disappears towards lower
frequencies. The figures for the second component shows the expected
selection against strong and weak pulses at low and high frequencies.

We have to conclude from the above that there is a group of strong
pulses occurring mostly at the longitudes of the first component,
which appear to be much stronger than the average pulses.

\begin{table}
\caption{Spectral indices obtained for PSR B1133+16 
in the frequency range from 340 to 4850 MHz. The result
using the published flux densities is quoted in Col.~2,
using the flux densities from the average profile in Col.~3
and the mean value of the single pulse spectra in Col.~4.
Corresponding medians are quoted in brackets.
\label{spec1133tab}}

\tabcolsep3pt
\begin{tabular}{lcccc}
\hline
\hline
 & Published & Average & \multicolumn{2}{c}{Single} \\
\hline
Full pulse & $-1.89\pm0.07$ & $-1.8\pm0.1$ & $-2.0\pm0.1$ & ($-2.08$)\\
Comp.~1 & -- &  $-1.7\pm0.1$ & $-2.0\pm0.1$ & ($-2.02$) \\
Comp.~2 & -- &  $-2.1\pm0.1$ & $-2.1\pm0.1$ & ($-2.23$)\\
\hline
\end{tabular}
\end{table}

\begin{figure}[t]

%\resizebox{\hsize}{!}{\includegraphics{spec1133c1.ps}}
%\resizebox{\hsize}{!}{\includegraphics{spec1133c2.ps}}

\psfig{file=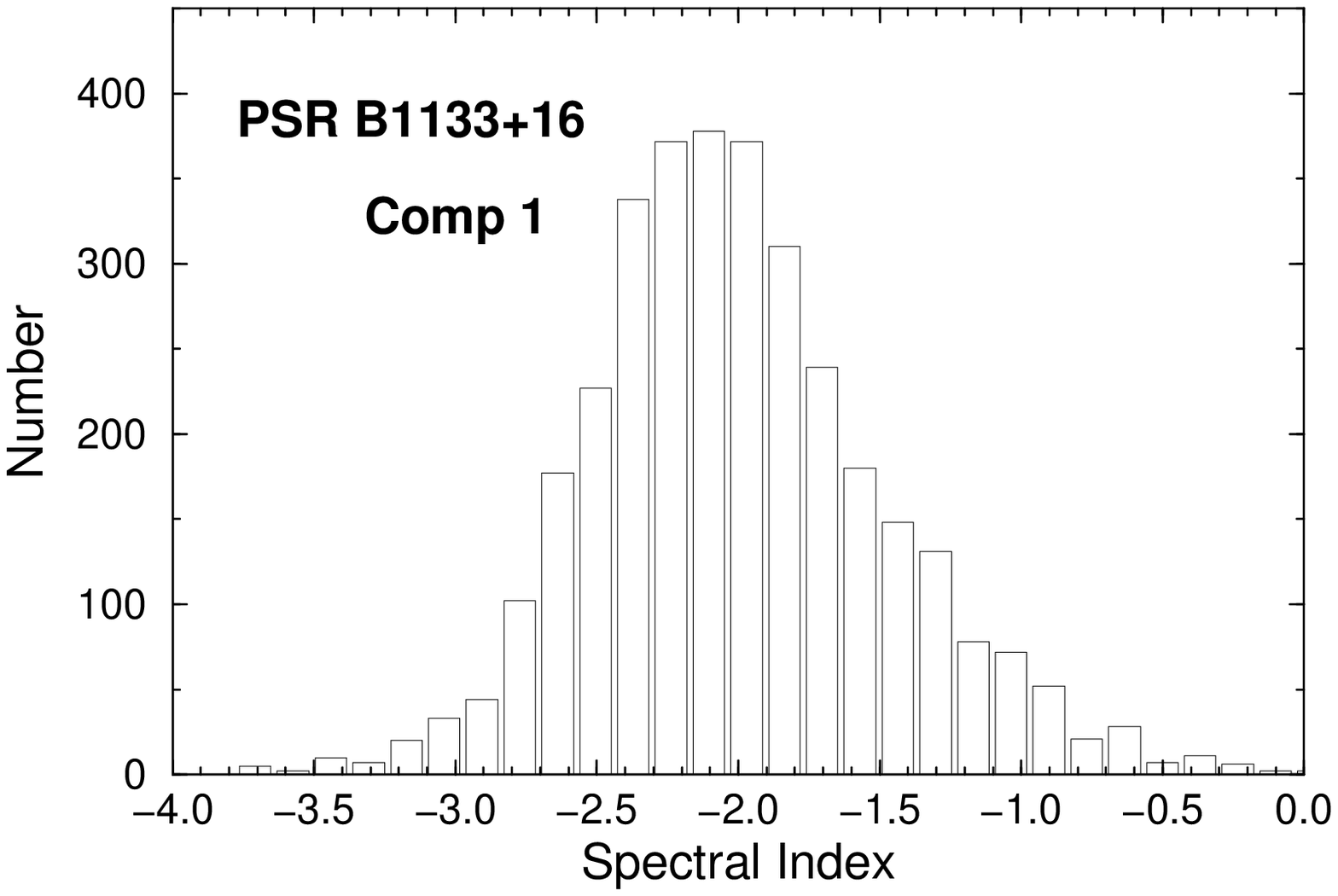,width=8cm}
\psfig{file=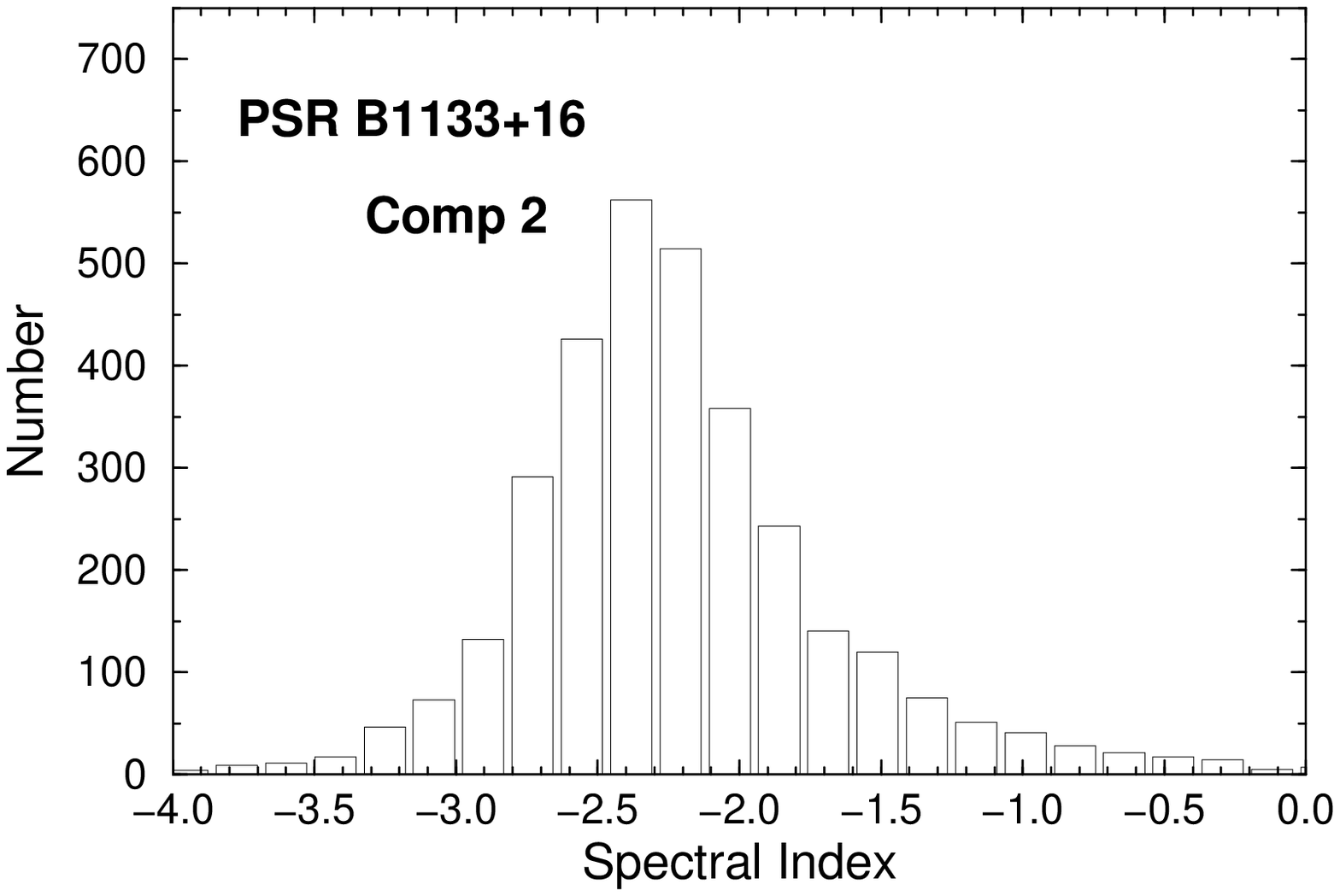,width=8cm}

\caption{\label{spec1133cfig}
Distribution of spectral indices obtained for single pulses
of PSR B1133+16 using the two defined pulse windows.
}
\end{figure}

\subsection{``Giant pulses'' in PSR B1133+16 ?}

\begin{figure}[t]
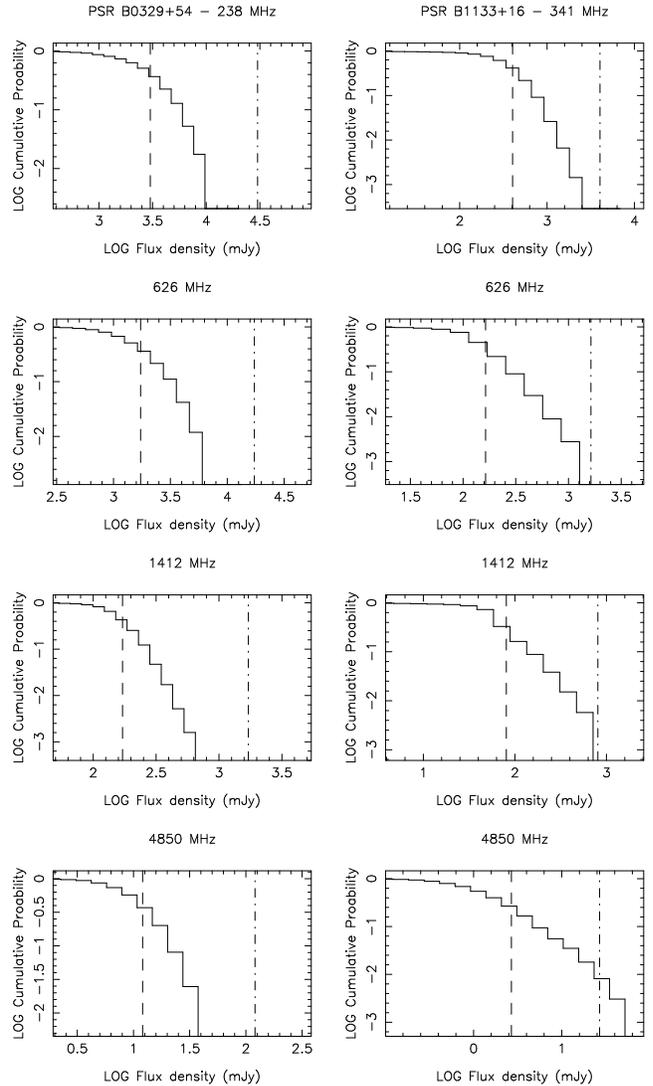


\begin{tabular}{cc}
\psfig{file=H4355F16A.ps,width=4cm} &
\psfig{file=H4355F16B.ps,width=4cm} 
\end{tabular}

\caption{\label{fluxdist0329}\label{fluxdist1133}
Cumulative probability
function  of the single pulse equivalent continuum
flux densities of
PSR B0329+54 (left) and PSR B1133+16 (right). Frequencies
238 MHz, 626 MHz, 1412 MHz and 4850 MHz are shown from top
to bottom. The dashed lines 
mark the mean of the equivalent continuum 
 flux density at the given frequency, 
while the dot-dashed lines indicate 10 times the mean.
}
\end{figure}

In order to study the occurrence of strong pulses for PSR B1133+16 and
to compare this to PSR B0329+54, we compute the cumulative probability
distribution of the single pulse flux densities, shown in
Fig.~\ref{fluxdist0329} for all frequencies.  Note that for PSR
B1133+16 we again only include detected (i.e.~non-nulling) pulses.

In these plots we indicate the mean flux density, $<\!S\!>$, measured at
the given frequency by a dashed line and also show a dotted-dashed
line marking a value of $10\times\!<\!S\!>$.  While the single pulses of
PSR B0329+54 are well below this threshold at all frequencies
(i.e.~no pulse is stronger than even 5 times the mean at any
frequency), for PSR B1133+16 we observe a number of pulses at 4850 MHz
which well exceed the corresponding value. We also observe a shift in
the location of the $10\times<\!S\!>$-line from low to high frequencies
relative to the cut-off of the distributions.  Therefore, in contrast
to PSR B0329+54, PSR B1133+16 has a significant fraction of strong
pulses which appear to become more frequent towards higher
frequencies.

In total we identify 40 pulses of PSR B1133+16 at 4850 MHz, whose mean
flux density exceeds 10 times the mean. This threshold is commonly
used to define so-called ``giant pulses'' which have been observed for
the Crab pulsar, PSR B1937+21 and PSR B1821$-$24 (see Johnston \&
Romani 2002 \nocite{jr02} for a recent review). In addition to their
large mean flux density, giant pulses also show a power-law
distribution in their energy distributions and tend to occur at a
specific range of pulse phases.  Giant pulses also appear to be
broadband by exceeding the mean flux density by a factor of 10 or more
at all frequencies.  The latter point is mostly based on
non-simultaneous observations by studying the energy distribution at
different frequencies (e.g.~Lundgren et al.~1995), but simultaneous
dual-frequency observations have been made for the Crab pulsar (see
Sallmen et al.~1999 and references therein) where only 70\% of all
giant pulses are seen both at 600 and 1400 MHz. The results of these
studies for the spectral index of giant pulses are somewhat
inconclusive.  While earlier observations conclude that, on average,
the spectral index of giant pulses is flatter than the average main
pulse spectral index, this finding is not supported by the most recent
study where the average spectral index of the giant pulses is
comparable to that of the average main pulse (see discussion by
Sallmen et al.~1999).  However, the Crab pulsar has a very steep radio
spectrum, so little may be learnt from the relative strength of its
normal and giant pulses at either extreme end of the radio spectrum.
However, we can compare other aspects of the strong pulses from PSR
B1133+16 pulses to the known properties of giant pulses or those of
so-called ``giant micropulses''. The latter have been discovered for the
Vela pulsar as very strong narrow pulses of very small width occurring
at fixed, narrow phases (Johnston et al.~2001), also showing a
power-law energy distribution.

While a power-law visible in the cumulative probability function of
giant pulse flux densities clearly separates them from normal pulses,
this appears not to be the case for PSR B1133+16 in
Fig.~\ref{fluxdist1133}.  It is possible that a number of 40
``giant'' pulses is too small to produce a recognisable feature, but a
rate of one strong pulse in every 100 (at 4850 MHz) is a much higher
rate than for PSRs B1821$-$24 and B1937+21 and is comparable to that
of the Crab pulsar. The noticeable change in slope of the
high energy end of PSR B1133+16's cumulative probability functions,
when going from low to high radio frequencies, may be an indication of
an emerging power-law component.

\begin{figure}[t]

\psfig{file=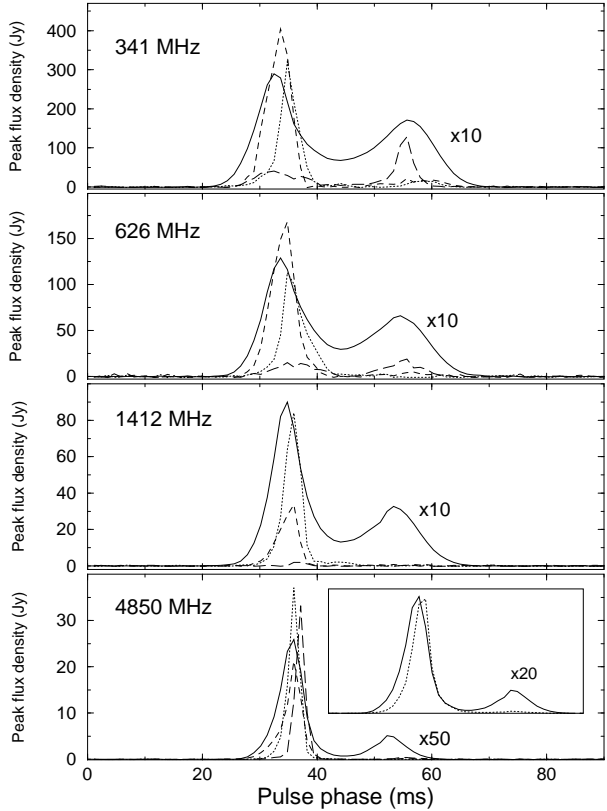,width=8cm}

\caption{\label{giants}
The three strongest pulses of PSR B1133+16
observed at 4850 MHz, showing simultaneously
their occurrence at the lower frequencies. The average pulse profile
(solid line) at these frequencies is also shown, multiplied by 50
at 4850 MHz and by 10 at the other frequencies. 
The y-axis shows corresponding peak flux densities.
The inset in the
lowest plot shows the average profile computed from the 40 ``giant''
pulses at 4850 MHz relative to the average pulse profile, multiplied
by a factor of 20. One can see that the strong pulses tend to occur
at the trailing edge of the leading component, being slightly
offset from its centre.}
\end{figure}

\begin{figure}[t]

\psfig{file=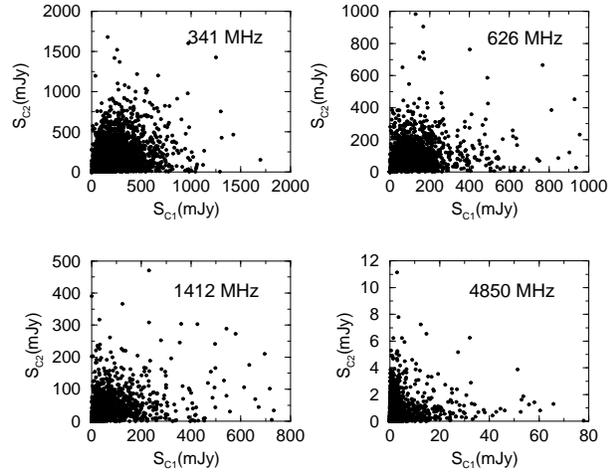,width=8cm}

\caption{\label{correlc1c2}
Correlation of the equivalent continuum
flux density of both components, C1 and C2, 
in PSR B1133+16 for all observed frequencies. While the correlation
is generally weak, the components appear to be independent at the
highest frequency.
}
\end{figure}

It appears that the strong pulses in PSR B1133+16 preferably occur at
the pulse phase of the leading component. This is verified by
computing the average profile from the 40 strong pulses at 4850
MHz. This profile is shown as the inset to Fig.~\ref{giants} where
we compare it to the average pulse profile. It becomes obvious that
the strong pulses appear to be narrower and indeed appear mostly at
the trailing edge of the leading component, being slightly offset from
its centre. This implies that any correlation in the flux densities of
the two normal profile components should become weaker at higher
frequencies, when the strong pulses in the leading component clearly
dominate the energetics. That is indeed the case (see
Fig.~\ref{correlc1c2}) where the flux densities appear almost
anti-correlated at 4850 MHz. This observation also explains the fact
mentioned in Sect.~4.1 that the correlation coefficient between
frequency pairs are larger for each component separately than for the
full profile.

We can study the broadband characteristics for PSR B1133+16 by
inspecting our 40 strong pulses at the lower frequencies. The three
strongest pulses are shown in Fig.~\ref{giants}. It turns out that
36 out of the 40 pulses are also much stronger than the average pulse
at a given phase at the lower frequencies. But since the emission of
these pulses is mostly concentrated in longitudes of the leading
component, while the second component is significantly stronger at the
lower frequencies, the mean flux density does not exceed the
giant-pulse threshold at those. There are 4 pulses (or 10\%),
however, which do not appear to be broadband in their strength.  One
of those is shown in the same Figure.

Based on the very small sample of pulsars known to exhibit giant
pulses, it had been suggested that the occurrence of giant pulses may
be related to the magnetic field strength at the light cylinder
(e.g.~Cognard et al.~1996) or to the emission process creating X-ray
and gamma-ray emission (Johnston \& Romani 2002). While PSR B1133+16
is not among the list of detected X-ray emitters, its magnetic field
at the light cylinder is five orders of magnitude smaller than that
of the Crab pulsar or PSR B1937+21.  Overall, however, it seems
possible that the strong pulses seen for PSR B1133+16 at 4850 MHz are
a population of giant pulses which only manifest themselves at the
higher frequencies because their radio spectrum is significantly
flatter than that of the normal pulsar emission. This is consistent
with observations of single pulses of PSR B1133+16 at an even higher
frequency of 8450 MHz where the ``giant'' pulses appear even more
frequently (Maron \& L\"ohmer, private communication).  Certainly, a
classification as giant pulses cannot be based on the measured flux
density alone (see also Johnston \& Romani 2002), so that better
statistics are needed to determine as to whether a power-law is
present in their energy distribution to confirm their nature.

\section{Summary \& Conclusion}\label{summary}

Analysing simultaneous multi-frequency observations of PSRs B0329+54
and B1133+16 we have derived intrinsic flux density time series by
correcting for scintillation effects. The resulting pulse-to-pulse
modulation indices show a minimum around 1 GHz and increase above and
below this frequency. This effect may be caused by a loss in
coherence. Indeed, inspecting the correlations between the flux
densities of different frequency pairs, it is striking that the
correlation decreases as the frequency widens. Similar effects have
been observed for the polarization properties in Karastergiou et
al.~(2001, 2002) and are clear indications for an intrinsic bandwidth
of the emission process or propagation effects in the magnetosphere.

We determined the intrinsic flux density spectra for single pulses of
PSR B0329+54 and derive similar conclusions as obtained from the
average profiles. This suggest a relatively stable formation process
of the average pulse profile, with spectral indices which are flatter
for the central part than for the outer components. The distributions
of spectral indices are wider for the outer conal component which are
comparable in width to those of PSR B1133+16. This may be also related
to the observation of weaker correlations observed for outer
components of PSR B0329+54 and could be linked to a line-of-sight
further away from the magnetic meridian.  The
flux densities below the spectral low-frequency turn-over seem to be
still related to the spectrum and flux densities at the higher
frequencies.

The spectral index distributions for the single pulses of PSR B1133+16
show significant deviations from a Gaussian shape. This asymmetry is
caused by strong pulses which qualify as giant pulses according to
their mean flux density at 4850 MHz. We have shown that most of these
pulses are broadband and occur at a relatively narrow range of phases
in the trailing edge of the leading component. All these properties
suggest that these pulses may be indeed related to the giant-pulse
phenomenon, whereas their relative spectral index is such that they
only become dominant at the higher frequencies. However, a power-law
behaviour in the energy distributions has not been observed yet and
better statistics is required. Hence, the main conclusions of this
paper are:
\begin{itemize}
\item[-] The intrinsic pulse-to-pulse modulation has a minimum
 around 1 GHz.
\item[-] The correlations in flux densities between frequency 
pairs becomes weaker as the frequency separation widens. The
best correlation is obtain between 600 and 1400 MHz, and it
becomes worse for outer profile components.
\item[-] The distribution of spectral indices measured for single pulses
of PSR B1133+16 is asymmetric (non-Gaussian), caused by a population
of strong, flat-spectrum  pulses.
\item[-] These strong pulses show a number of features which suggests
their nature as giant pulses.
\end{itemize}
We have demonstrated that the study of the radio spectra
of single pulses represents a so-far untapped reservoir of information
directly related to the radiating particles and their emission
process. Understanding their properties (unclouded by any averaging
process) appears to be essential in the identification of the
underlying emission process.

\acknowledgements We are very grateful to Christine Jordan and Axel
Jessner for help with the observations and data reduction and
acknowledge fruitful discussions with Graham Smith. It is a pleasure
to thank the referee Misha Popov for a careful reading and his
constructive comments and suggestions. We thank Maura
McLaughlin for comments on the manuscript.

\end{document}